# STATISTICAL MODELING FOR WIND-TEMPERATURE METEOROLOGICAL ELEMENTS IN TROPOSPHERE

A. Virtser[1], Yu. Shtemler[2], E. Golbraikh[3]

[1]Institute for Industrial Mathematics, Beer-Sheva, 84311, Israel

[2]Department of Mechanical Engineering,

[3]Department of Physics,

Ben-Gurion University of the Negev, P.O.B. 653, Beer-Sheva 84105, Israel

**Abstract**

A comprehensive statistical model for vertical profiles of the horizontal wind and temperature throughout the troposphere is presented. The model is based on radiosonde measurements of wind and temperature during several years. The profiles measured under quite different atmospheric conditions exhibit qualitative similarity, and a proper choice of the reference scales for the wind, temperature and altitude levels allows to consider the measurement data as realizations of a random process with universal characteristics: means, the basic functions and parameters of standard distributions for transform coefficients of the Principal Component Analysis. The features of the atmospheric conditions are described by statistical characteristics of the wind-temperature ensemble of dimensional reference scales. The high effectiveness of the proposed approach is provided by a similarity of wind – temperature vertical profiles, which allow to carry out the statistical modeling in the low-dimension space of the dimensional reference scales for the wind, temperature and altitude levels. The knowledge of the joint wind-temperature distributions in altitude open wide perspectives for modeling of various physical processes in real atmospheric conditions like the transfer of passive scalars or impurities, the electromagnetic wave penetration, the instability of internal waves, transition to turbulence etc.

## 1. Introduction

Research and modeling of various physical processes occurring in Earth's troposphere like transfer of a passive scalars or impurities, penetration of the electromagnetic wave, generation of the inertio- gravity internal waves, transition to turbulence etc. requires knowledge of the distribution of its basic parameters, such as wind speed and temperature. Such an approach allows to employ the wind speed and temperature in real atmospheric conditions instead of frequently used models for sheared wind and temperature vertical profiles like tanh- or log- functions of the altitude (e.g. Whiteway et al. 2004). Preliminary information on turbulence as well as on troposheric aerosols during different atmospheric conditions based on the Richardon number can be infered from the measured wind-temperature data (Scinocca and Ford, 2000, Wilson 2004, Aristidi et al. 2005, Dutta et al. 2009, Castracane et al. 2001, Zilberman et al., 2008).

The Dryden and von Karman models and their developments are widely used in aviation design for wind turbulence modeling (see, e.g., *Military specifications* 1974). In advanced models for probability distribution of wind, the measurement data are subdivided into meteorologically uniform portions corresponding to the wind coming from a sector of directions during a time period of the year. After that the Weibull model is applied individually to each portion. Parameters of the advanced model should be fitted to histograms of measurement data (see *Modelling of Atmospheric Flow Fields* 1996 and references therein). The input parameters of the models are properly defined within atmospheric boundary layer, while they are poorly defined at high altitudes. However, it is often necessary to know not only the average distributions of these variables, but their variations according to atmospheric conditions. Therefore, in this study, we continued to develop the approach proposed in (Virtser et al., 2005; Zilberman et al., 2008) concern with statistical processing of radiosonde measurements. In contrast to traditional wind - temperature statistical models at a fixed altitude, the approach is concerned with wind - temperature statistical modeling of vertical meteorological profiles throughout the troposphere (Virtser et al., 2005). In opposite to traditional models for the winds at a fixed altitude level, the present work is concerned with statistical modeling of vertical meteorological profiles throughout the troposphere. Winds are commonly treated as horizontal flows because maximal horizontal winds are typically in tens of meters per second, while vertical winds are of the order of one meter per second and can reach order of the horizontal winds only in emergencies such as thunderstorms. The present study is based on the similarity and dimensionality for the homogenization of meteorological data measured under different atmospheric conditions for winds exceeding 3–5 m/s. A proper choice of reference scales for the scaling procedure makes the measurement data meteorologically homogeneous and dimensionless. The universality of the homogenized data is the key idea of the present approach. After that they are considered as realizations of a random process with universal profiles, i.e. of independent atmospheric conditions. To develop a model of mean wind and temperature, we have used meteorological data on troposphere parameters obtained over the Israeli territory during 1997–1999 (Virtser et al., 2005).

Statistical processing of meteorological data is complicated by their heterogeneity because of difference in atmospheric conditions under which they were measured (reference wind speed, directions, temperatures and altitudes). Thus, wind and temperature are strongly subjected to seasonal variations. The horizontal winds can be subdivided into strong and weak winds. If the troposphere is unstably stratified as it occurs in measurement data considered in the present study, different kind instabilities of the mean shear flow can occur: both atmospheric convection and turbulence typical for strong winds, or convection predominating under weak winds. The problem addressed here is development of a joint statistical model for the meteorological element consisting of the local horizontal wind and temperature profiles, while the data for relative humidity is beyond our interest. The present study is based on similarity and dimensionality grounds for homogenization of meteorological data measured under different atmospheric conditions for strong winds. A proper choice of the reference scales for a scaling procedure makes the measurement data meteorologically homogeneous and dimensionless, after that they are considered as realizations of a random process with universal, i.e. independent of atmospheric conditions. The features of the atmospheric conditions are taken into account by statistical characteristics of the joint wind-temperature ensemble of dimensional reference scales. The universality of the homogenized data is the key idea of the present approach. Furthermore, two hypotheses are adopted and tested in the current study. The former hypothesis adopted is that three ensembles are mutually independent: (i) joint wind-temperature ensemble of dimensional reference scales, two ensembles of the Principal Component Analysis (PCA) [see Attachment in Zilberman et al., 2008] transform coefficients for (ii) wind and (iii) temperature. The second hypothesis is the normal law for each of the three ensembles.

**2. Qualitative Description of Meteorological Data. Similarity of Winds and Temperatures**

Let us qualitatively describe the meteorological data gathered over a 3-year period and determine the dimensional reference scales for homogenization of the data measured under different atmospheric conditions for the vertical profiles of local horizontal wind $W = \{U, V\}$ and temperature $T$ depending on altitude $H$. Two-dimensional vector of horizontal wind $W$ is also described by the wind speed $|W|=W$ and direction $\alpha$ that the wind is blowing from. Unlike the common use in meteorology the wind direction $\alpha$ is measured counter-clockwise from the north as positive direction ($-180° \leq \alpha \leq 180°$, so the jumps of wind direction $\alpha$ can occur at $\alpha = \pm 180°$).

Although the wind data are discrete, for the convenience of description we operate below with continuous temperature $T_m(H)$, wind speed $W_m(H)$ and direction $\alpha_m(H)$ ($m=1, 2, ..., M$, $M$ is the total number of radiosonde measurements). The horizontal winds can be subdivided into strong and weak winds (see Fig. 1, where the hodograph, speed and direction are presented of the wind velocity $\mathbf{W}_m = \{U_m, V_m\}$). Figure 1 exhibits the following qualitative differences between the weak and strong winds:

- (a) the altitude level of the maximum wind speed is highly expressed for strong winds unlike to the weak winds;
- **(b)** the maximal $V_m$ component is significantly larger than the maximal $U_m$ component for sufficiently strong winds ($|V_m| < 85\, knot$, $|U_m| < 25\, knot|$ in Fig. 1 (a), 1 $knot$ =0.51 $m/s$), while for the weak wind both components are of the same order ($|V_m| \sim |U_m| < 20\, knot$ in Fig. 1 (e));
- **(c)** weak-wind direction $\alpha$ is widely varied with altitudes (Fig. 1 (d)), while in the strong-wind case the value of $\alpha$ can be estimated on average by its value $\alpha_m^{max}$ at the altitude level of the maximum wind. This is clearly illustrated by the straight-line OO' passing through two points in Fig. 1 (a) for the wind hodograph - of the sonde start (at the origin of the reference frame) and of the maximum wind speed.

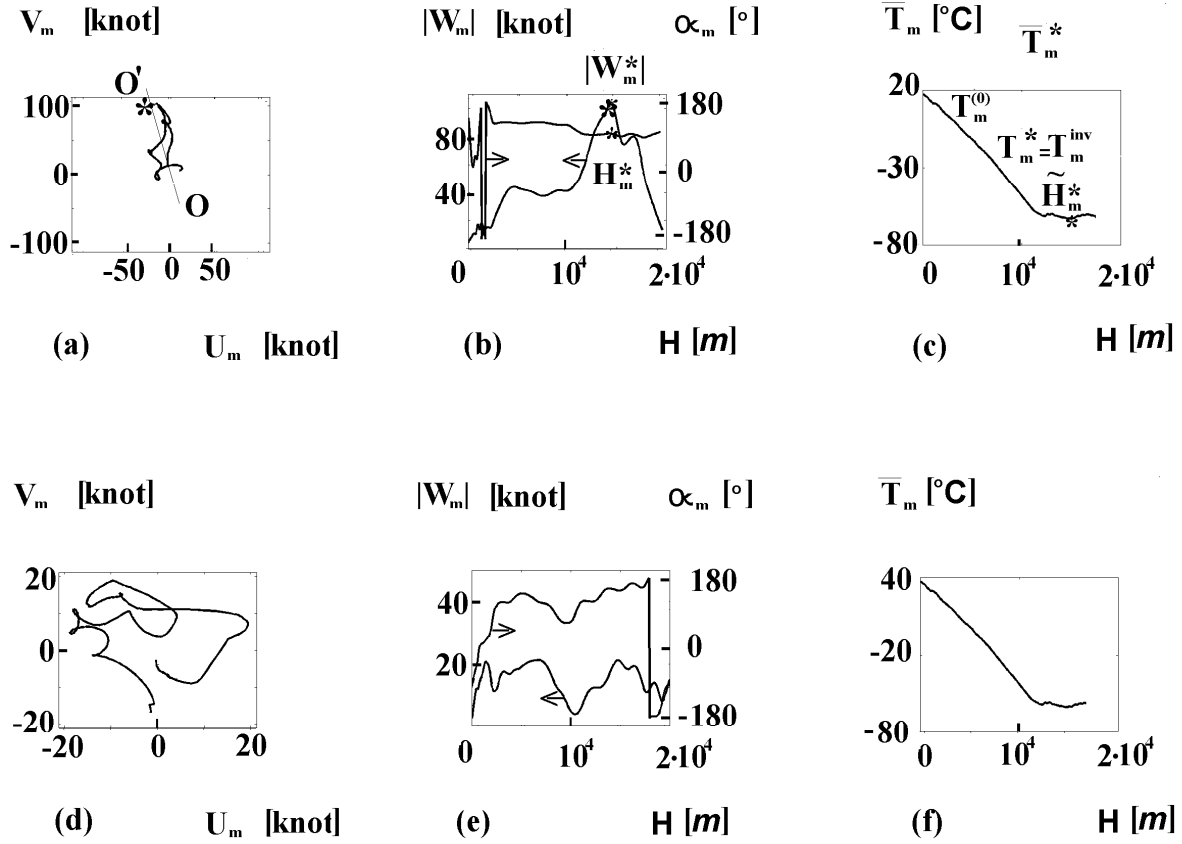

Figure 1. Typical wind holographs, vertical profiles of wind speed, direction and temperature for strong (a),(b), (c) and weak (d), (e), (f) winds.

\* point of the maximum wind speed or of the inverse temperature.

The strong winds measured at different times of the year under quite different atmospheric conditions exhibit qualitative similarity, but differ in the reference wind speed, direction and altitude level. Strong wind speed $W_m$ for a sonde $m$ rises from zero at the ground up to maximum value $W_m^{max}$ at altitude level $H_m^{max}$, $\alpha = \alpha_m^{max}$ and then drops with altitudes $H$ (Fig. 1 (b) ). Thus, it is naturally to define the following reference scales for strong winds:

$$W_m^* = W_m^{max}, \quad \alpha_m^* = \alpha_m^{max} \quad \text{and} \quad H_m^* = H_m^{max}.$$

In Fig. 2 distributions are presented (a) of the reference-wind (a) speed $|\mathbf{W}_m^*| \equiv W_m^*$ and (b) direction $\alpha_m^*$ vs. the reference altitude level $H_m^*$. Artificial clusters in the left and right lower corners in Fig. 2 (b) correspond to the jumps of the wind direction at $\alpha = \pm 180°$. The altitude level of the reference wind is varied about the mean value $\overline{H}_m^* \approx 12 \cdot 10^3 \, m$ for both weak and strong winds. However, Fig. 2 (a) demonstrates separation of the reference-wind data on two clusters of the strong and weak winds. The wind is strong if the maximum of the wind speed exceeds 50 *knots* and varied about the mean value with a relatively low dispersion. Note that about 70% of all winds observed during three years are strong. As was mentioned above, the altitude level of the maximum wind speed is highly expressed for sufficiently strong winds.

Since dry air almost fails to absorb solar radiation, the temperature of air is determined by its heat-exchange with the ground surface. In the present meteorological data the air temperature normally decreases with distance away from the warm ground surface up to the inversion of temperature. The inversion altitude may be considered as the width of the convective boundary layer, since it imposes a barrier to the heat transfer from the ground to the upper-lying atmosphere (*Modeling of Atmospheric Flow Fields* 1996). The characteristic vertical profile of temperature in Fig. 1 (c), (e) has the near-maximum value $T_m^{(0)}$ at zero level *H*=0, decreases up to inverse value $T_m^{inv}$ at altitude $H_m^{inv}$, and then rises with altitude. Measuring the temperature from the referring level $T_m^{(0)}$, we define the following reference scales for the temperature:

$$T_m^* = | T_m^{inv} - T_m^{(0)} | \text{ and } \widetilde{H}_m^* = H_m^{inv}.$$

In Fig. 2 (c) the distribution of the reference temperature $T_m^*$ with the altitude level $\widetilde{H}_m^*$ is shown The altitude level of the inverse temperature equals $\overline{\widetilde{H}}_m^* \approx 18 \cdot 10^3 \, m$ on average and exceeds the averaged reference altitude level for wind $\overline{H}_m^* \approx 12 \cdot 10^3 \, m$.

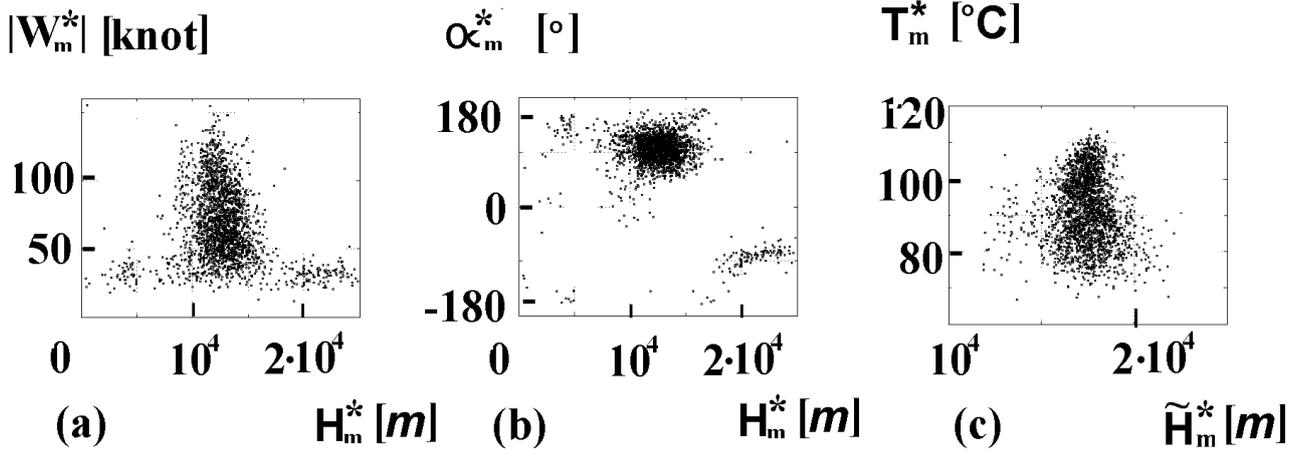

Figure 2. Reference scale distributions over the 3-year period.
Reference wind speed (a) and direction (b) vs. altitude level of maximal wind-speed.
Reference temperature (c) vs. altitude level of inverse temperature.

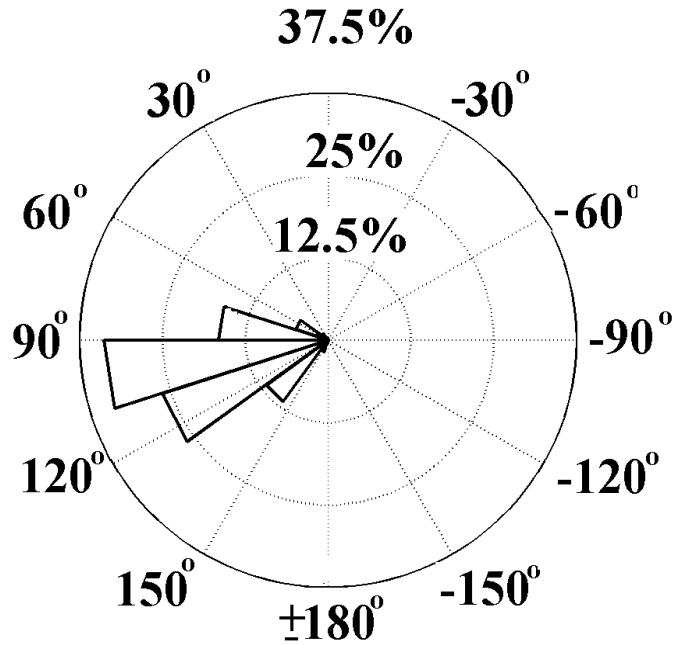

Figure 3. Maximal-wind rose for the strong winds over the 3-year period.

Note that the strong winds have the west-sound prevailing direction (about 85% of their total number) $70° < \alpha^{(0)} < 125°$ or $\alpha^{(0)} \approx 110°$ on average, as illustrated in Fig. 3 by the maximal-wind rose over 3 years. Hence $|\cos(\alpha_m^*)| \ll |\sin(\alpha_m^*)|$, and $|U_m^*| = |W_m^* \cos(\alpha_m^*)|$ is much less than $|V_m^*| = |W_m^* \sin(\alpha_m^*)|$ on average.

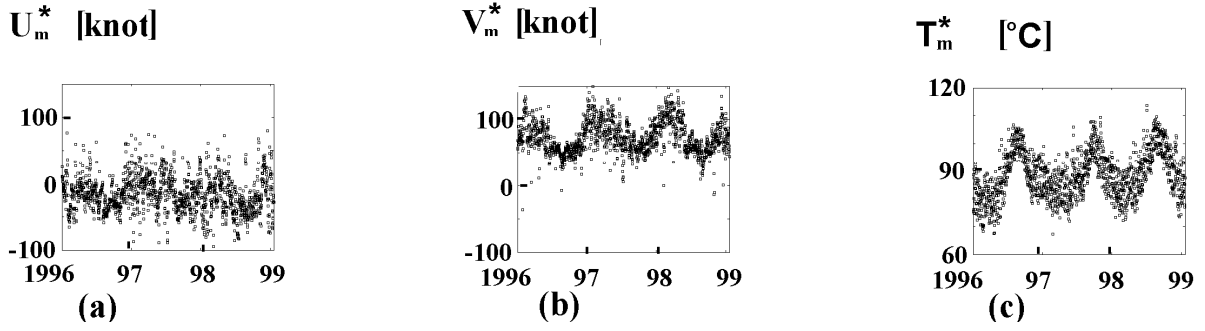

Figure 4. Reference scales (a) $U_m^*$, (b) $V_m^*$, (c) $T_m^*$ of meteorological elements vs date of measurement over the 3-year period.

In Fig. 4 the heterogeneity of atmospheric conditions is illustrated by seasonal variations of the reference scales over the 3-year period. Note that both day and evening time measurements are made daily, in particular, two ordinates correspond to every abscissa value in Fig. 4.

Below a scaling procedure is described excluding influence of the seasonal variations on the meteorological data by their homogenization.

## 3. Meteorological Data: Homogenization, Means and Deviations, Statistical Properties

Let us consider a homogenization procedure based on the reference scales of the wind and temperature determined above.

*Homogenization of the Wind and Temperature Data*

Let us pass to the dimensionless variables by scaling wind vector $\mathbf{W}_m$ by $W_m^*$ and rotating it by angle $\alpha_m^*$

$$\mathbf{w}_m(h) = \Omega(\alpha_m^*)\frac{\mathbf{W}_m(H)}{W_m^*}, \quad h = \frac{H}{H_m^*}, \tag{3.1}$$

where $m=1,2,\ldots M$, $\Omega(\alpha_m^*)$ - rotation matrix, $W_m^* = |\mathbf{W}_m^*|$ and $\alpha_m^*$ are speed and direction of reference wind $\mathbf{W}_m^* = \{U_m^*, V_m^*\}$, and $H_m^*$ - reference altitude level:

$$\mathbf{W}_m^* = \mathbf{W}_m^{\max}, \quad \Omega(\alpha_m^*) = \begin{pmatrix} \cos\alpha_m^* & \sin\alpha_m^* \\ -\sin\alpha_m^* & \cos\alpha_m^* \end{pmatrix}, \quad \alpha_m^* = \alpha_m^{\max}, \quad H_m^* = H_m^{\max}. \quad (3.2)$$

$$\mathbf{w}_m(h) = \frac{\mathbf{W}_m}{\mathbf{W}_m^*}, \quad h = \frac{H}{H_m^*}, \quad m=1,2,\ldots M.$$

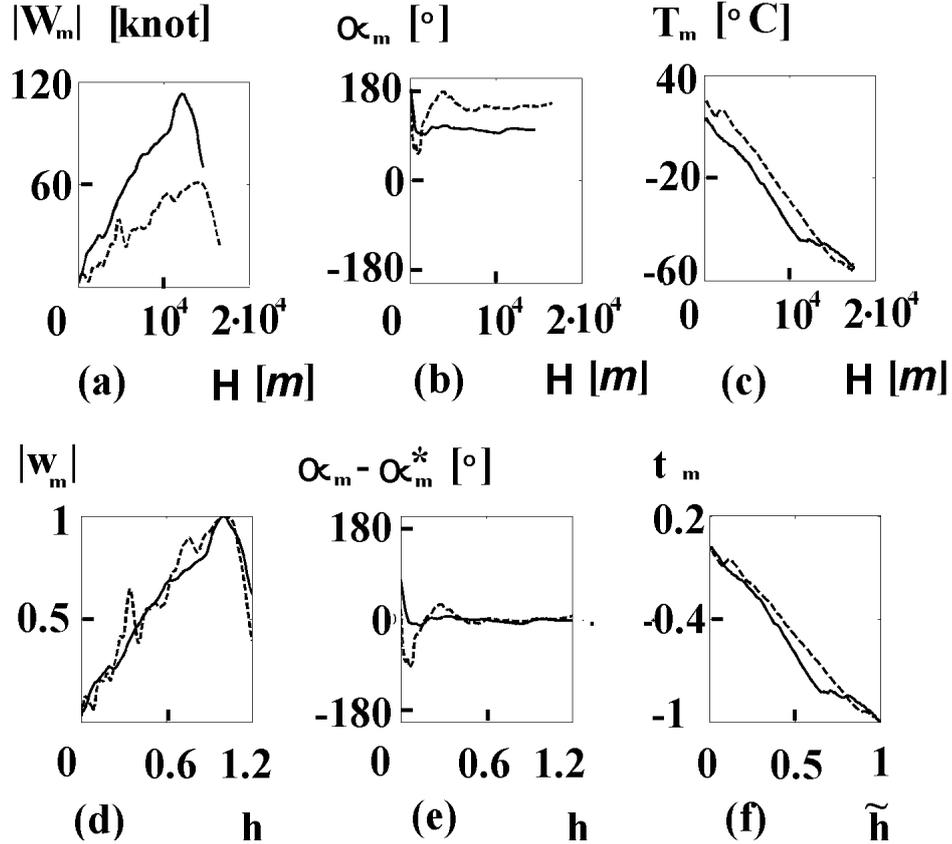

Figure 5. Two vertical profiles of wind speed, direction and temperature before (a),(b), (c) and after (d), (e), (f) homogenization (strong winds).

Wind speed $W_m^*$, direction $\alpha_m^*$ and altitude level of the maximum wind speed are determined directly from measurement data. To make dimensionless variables meteorologically homogeneous the wind vector in (3.1) is rotated by the characteristic angle along with the standard normalization by wind module. Transform (3.1) for 2-dimensional vectors can be rewritten in the alternative form of the standard scaling procedure. Retaining the previous vector notation for corresponding complex values $\mathbf{W}_m^* = |\mathbf{W}_m^*|\exp(i\alpha_m^*)$, $\mathbf{W}_n = |\mathbf{W}_n|\exp(i\alpha_n)$ and $\mathbf{w}_m = |\mathbf{w}_m|\exp[i(\alpha_m - \alpha_m^*)]$, we rewrite (3.1) as follows:

Analogously, for temperature scaling let us pass to the following dimensionless variables:

$$t_m(\widetilde{h}) = \frac{T_m - T_m^{(0)}}{T_m^*}, \quad \widetilde{h} = \frac{H}{\widetilde{H}_m^*} \qquad (3.3)$$

with the following dimensional reference scales:

$$T_m^* = |T_m^{inv} - T_m^{(0)}| \text{ and } \widetilde{H}_m^* = H_m^{inv}, \quad T_m^{inv} = T_m(H_m^{inv}), T_m^{(0)} = T_m(0). \qquad (3.4)$$

The auxiliary reference frame is also defined as rotated by the angle $\alpha_m^*$ equal to the mean direction of the wind profile so that the *x* axis will be aligned with the reference wind vector. The wind components in the auxiliary reference frame will be referred to as lateral and transversal $\hat{\mathbf{W}}_m = \{\hat{U}_m, \hat{V}_m\}$

$$\hat{\mathbf{W}}_m(H) = \Omega(\alpha_m^*)\mathbf{W}_m(H), \qquad m=1,2,\ldots M, \qquad (3.5)$$

The scaling results for strong winds are illustrated by two vertical profiles of wind speed, direction and temperature before and after homogenization (Fig. 5). Relatively large perturbations of wind directions occur at low levels (see Fig. 5 (e) ). Note in this connection that wind measurements at the near ground region are subjected to significant errors in establishing of the Global Positioning Satellites locations.

*Mean Wind and Temperature*

The mean wind and temperature can be introduced as follows:

$$\overline{\mathbf{w}}(h) = \frac{1}{M}\sum_{m=1}^{M}\mathbf{w}_m(h), \quad \overline{t}(\widetilde{h}) = \frac{1}{M}\sum_{m=1}^{M}t_m(\widetilde{h}). \qquad (3.6)$$

In Fig. 6 (a) and (b) the dimensionless mean speed and temperature $\overline{\mathbf{w}}(h) = \{\overline{u}, \overline{v}\}$ and $\overline{t}(\widetilde{h})$ are shown. Note that transversal component of wind is much less than longitudinal one, and the mean wind speed (approximately equal to the longitudinal component of wind) and the mean temperature are near-linearly varied with altitude at least lower than the altitude level of maximum wind. Note that analogous behavior is also typical for the mean potential temperature that indicates on the unstable stratification. As well known in such case the following shear flow instabilities are typical: atmospheric convection and turbulence.

*PCA Expansion for Wind Deviations*

Applying the standard procedure of Principal Component Analysis (e.g. Rencher 1998) to deviations from the mean wind

$$\mathbf{w}'_m(h) = \mathbf{w}_m(h) - \overline{\mathbf{w}}(h), \quad m = 1, 2, \ldots M, \tag{3.7}$$

one has the following expansion for each *m*:

$$\mathbf{w}'_m(h) = \left[ \sum_{k=1}^{2K} \mathbf{c}_m^{(k)} \mathbf{v}^{(k)}(h) \right] + \boldsymbol{\varepsilon}_m(h), \tag{3.8}$$

Here basis functions $\mathbf{v}^{(k)}(h)$ ($k = 1,2,\ldots,2K$) and error of the expansion $\varepsilon_m(h)$ are defined within interval $(0, \eta)$. The dimensionless parameter $\eta$ should be chosen from physical reasons under the condition that dimensional measurements $\mathbf{W}_m$ for all $m=1,2,\ldots, M$ are at least within the interval $(0, \eta H_m^{max})$, $\eta \geq 1$. The number of the basis functions $2K$ is assumed to be equals the number of the discrete points at any profile. The first ($k=1, 2$) and second ($k=3, 4$) modes $\mathbf{v}^{(k)}(h) = \{u^{(k)}, v^{(k)}\}$ are shown vs. *h* in Fig. 6 (c) and (d). The number of modes *K* is determined as equal to the number of extremum points of basic functions. In calculations, the consequent pairs of basic vector functions corresponding to the same mode number will be used simultaneously. Note that both the first ($k=1, 2$) and second ($k=3, 4$) modes have the following "symmetry":

$$u^{(k)} \sim v^{(k+1)}, \ v^{(k)} \sim -u^{(k+1)}, \ k = 1 \text{ and } 3,$$

where ~ denotes qualitative similarity. In calculations, the consequent pairs of basic vector functions corresponding to the same mode number will be used simultaneously. Basic functions $\mathbf{v}^{(k)}$ satisfy the eigenvalue problem (A1.5) $\mathbf{B}\mathbf{v} = \lambda \mathbf{v}$ with covariance operator $\mathbf{B}$ and eigenvalue $\lambda$ (Zilberman et al., 2008). In the present calculations we use $\eta=1.2$, $K=100$ with vertical grid with $2K$ cells and covariance matrix $\mathbf{B}$ with $2K*M$ elements.

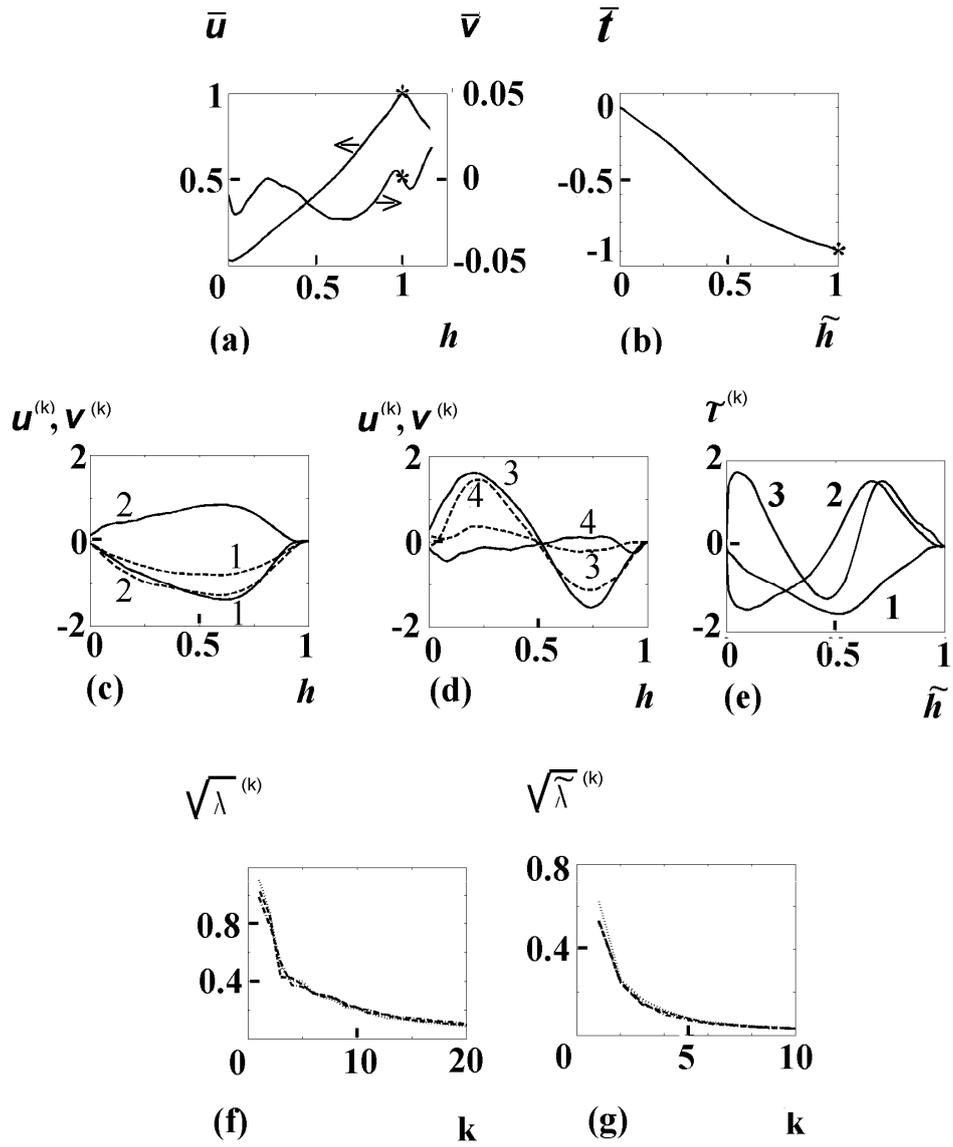

Figure 6. Dimensionless means, basic functions and standard deviations.

Mean (a) wind speed $\{\bar{u}, \bar{v}\}$ vs. $h$, (b) and mean temperature $\bar{t}$ vs. $\tilde{h}$,

* point of the wind speed maximum or of the inversion temperature.

Wind basic functions $\{u^{(k)}, v^{(k)}\}$ vs. $h$, (c) first mode ($k$=1, 2), (d) second mode ($k$=3, 4).

Temperature basic functions (e) $\tau^{(k)}$ vs. $\tilde{h}$ first three modes ($k$=1, 2, 3).

Standard deviations $\sqrt{\lambda^{(k)}}$ and $\sqrt{\tilde{\lambda}^{(\tilde{k})}}$ for (f) wind and (g) temperature vs. current number of mode.

Eigenvectors $\mathbf{v}(h) = \mathbf{v}^{(k)}(h)$ yield an optimal solution to the problem

$$\frac{1}{M}\sum_{m=1}^{M} \|\varepsilon_m\|^2 = \frac{1}{M}\sum_{m=1}^{M} \|\mathbf{w'_m} - \sum_{k=1}^{2K} c_m^{(k)} \mathbf{v}^{(k)}\|^2 \to \min$$

minimizing the mean-square error of the expansion. The wind mean-square error $S$ can be calculated as a dimensional function of the mode number $K$

$$S^2 = \frac{1}{M}\sum_{m=1}^{M} \|\mathbf{W}_m^*\|^2 \cdot \|\mathbf{w}_m(h) - \overline{\mathbf{w}}(h) - \sum_{k=1}^{2K} c_m^{(k)} \mathbf{v}^{(k)}(h)\|^2 \ . \tag{3.9}$$

*PCA Expansion for Temperature Deviations*

Analogously, for deviations from the mean temperature

$$t'_m(\widetilde{h}) = t_m(\widetilde{h}) - \bar{t}(\widetilde{h}), \qquad m=1,2,\ldots,M, \tag{3.10}$$

we have the following expansion for each $m$:

$$t'_m(\widetilde{h}) = [\sum_{k=1}^{K} \widetilde{c}_m^{(k)} \tau^{(k)}(\widetilde{h})] + \widetilde{\varepsilon}_m(\widetilde{h}) . \tag{3.11}$$

Here the upper tilde is used for the temperature variables. For convenience, the analogous notation is employed for the temperature variables as for the wind variables, but it is taken into account that the temperature is a scalar value in opposite to the vector of wind. The same number of modes in both wind and temperature expansions is assumed. In Fig. 6 (e) basic functions $\tau^{(k)}$ $k$=1, 2, 3 are shown vs. $\widetilde{h}$. Here temperature data are assumed to be known at least within interval $(0, \widetilde{\eta}\widetilde{H}_m^{\max})$. In the present calculations we set $K$=100, $\widetilde{\eta}$=1.01.

Analogously to (3.9) the wind mean-square error $\widetilde{S}$ for temperature can be calculated as a dimensional function of the number of basic functions $K$

$$\widetilde{S}^2 = \frac{1}{M}\sum_{m=1}^{M} \|\mathrm{T}_m^*\|^2 \cdot \|t_m(\widetilde{h}) - \bar{t}(\widetilde{h}) - \sum_{k=1}^{K} \widetilde{c}_m^{(k)} \tau^{(k)}(\widetilde{h})\|^2 \ . \tag{3.12}$$

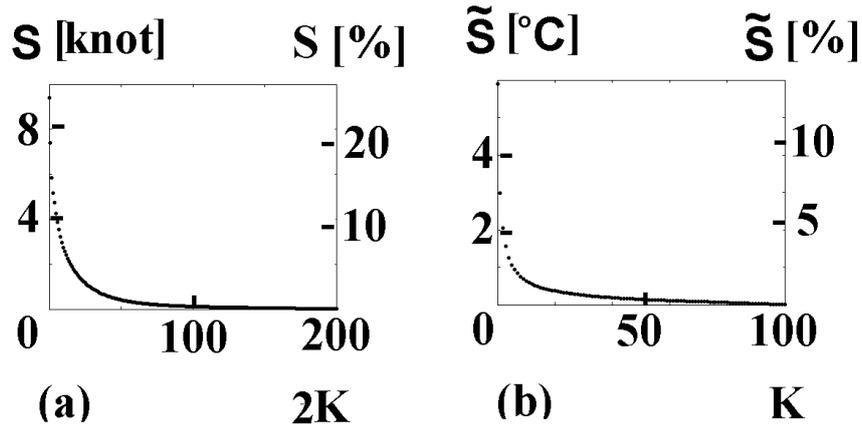

Figure 7. Dimensional and dimensionless mean-square errors for (a) wind and (b) temperature vs total number of basic functions $(K = 1,...,100)$.

Note that the value $K=0$ in the above relations correspond to approximations of the mean wind and temperature. In accordance with Fig. 7, dimensional mean-square errors for the mean wind and temperature are 9.5 *knots* and 5.9 °C ($K=0$), 3-mode approximation ($K=3$) yields 5 *knots* and 2 °C, and 5-mode approximation ($K=5$) monotonically reduces the dimensional mean-square errors to 2.5 *knots* and 1°C. In accordance with available experimental data the micro-scale components of wind and vertical winds that are beyond our consideration can be of the order of 2-5 *knots* (*Engineering Meteorology* 1982). So, although any required exactness can be achieved, too high exactness has little meaning, and physically admissible values of the wind mean-square error correspond to 5- or even 3-mode approximation (see Fig. 8 (a)-(c)). Note that at $K=3$, $K$-mode approximation corresponds to the wave length of the order of $\overline{H}_m^* / K \sim 4 \cdot 10^3$ *m* in the vertical direction.

*Statistical Properties of PCA Transform Coefficients*

It is known that all transform coefficients $c_m^{(k)}$ of Principal Component Analysis values are uncorrelated for different $k$, have zero means and variances equal to eigenvalues $\lambda^{(k)}$:

$$\frac{1}{M}\sum_{m=1}^{M} c_m^{(k)} c_m^{(n)} = 0 \ (n \neq k), \quad \frac{1}{M}\sum_{m=1}^{M} c_m^{(k)} = 0, \quad \frac{1}{M}\sum_{m=1}^{M} c_m^{(k)2} = \lambda^{(k)}. \tag{3.13}$$

The same is valid for the temperature coefficients $\tilde{c}_m^{(k)}$, but with variances $\tilde{\lambda}_m^{(k)}$.

*Annual Stability of Statistical Characteristics of the Local Meteorology*

The dimensionless mean wind was calculated independently for three consequent years to analyze its dependence on the year of measurements. It is seen that the dimensionless mean wind and temperature (Fig. 9 (a-b)), PCA basic functions (Figs. 9 (c-d) and standard deviations (Fig. 9 (e-f)) depend weakly on the year of measurement for three years.

Figure 9 exhibits the annual stability of the meteorological database at least for the main statistical characteristics of the low order modes. The annual characteristics of the higher order modes, however, were subject to significantly larger variations from year to year.

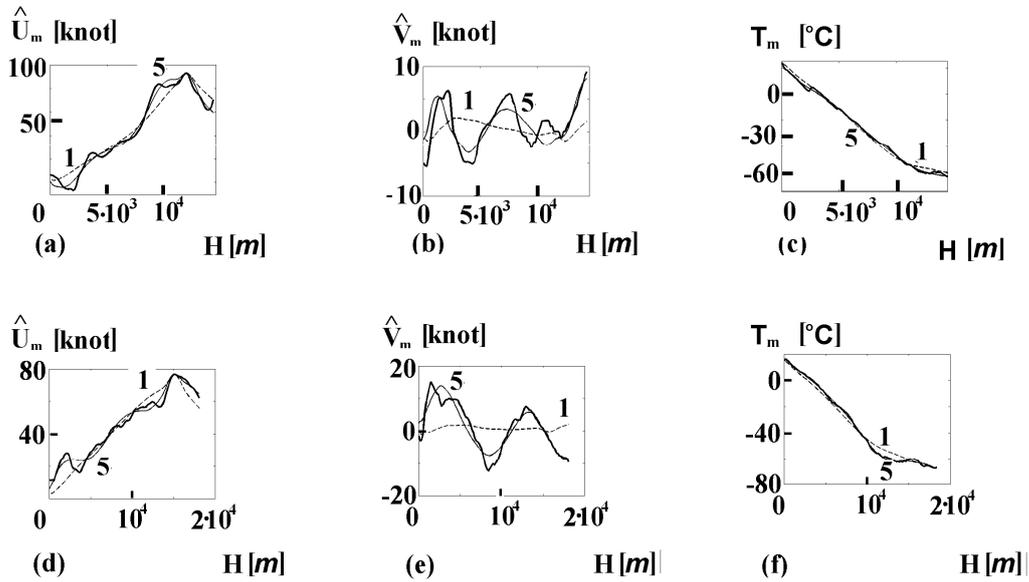

Figure 8. Measured (a-c) and simulated (d-f) vertical profiles (bold curves) and their PCA approximations at $K= 1$ and $5$ (thin and dash curves) at reference angle $\alpha_m^*=70^o$. Longitudinal (a), (d) and transversal (b), (e) components of wind and temperature (c), (f)

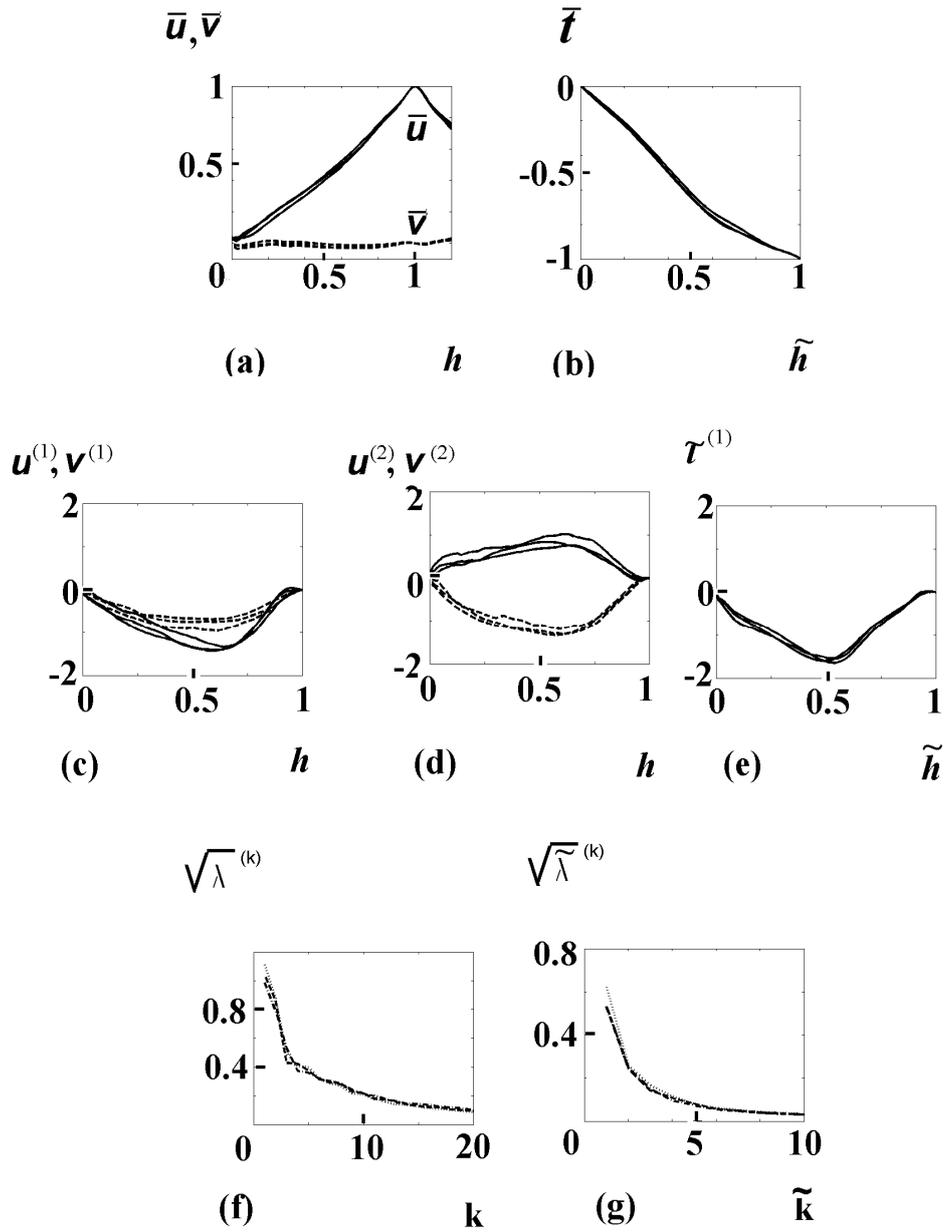

Figure 9. Dimensionless statistical characteristics for 3-year period.
Dimensionless mean (a) wind $\overline{\mathbf{w}}(h)=\{\overline{u},\overline{v}\}$ vs. $h$, (b) temperature $\overline{t}(\tilde{h})$ vs $\tilde{h}$,
First-mode dimensionless basic functions for wind (c), (d) and temperature (e),
$\mathbf{v}^{(k)}(h)=\{u^{(k)},v^{(k)}\}$ $k=1, 2$ vs. $h$, $\overline{u}$, $u^{(k)}$ - solid lines, $\overline{v}$, $v^{(k)}$ - dashed lines.
Dimensionless standard deviations for (f) wind and (g) temperature vs current number of basic functions.

## 4. Statistical Modeling and Simulations of Local Meteorology

For statistical modeling of local meteorology the following scheme is proposed. According to transforms (3.1), (3.3) and expansions (3.8), (3.11), *K*-approximation of wind $\mathbf{W}(H)$ and temperature $T(H)$ are presented in the form:

$$\mathbf{W}(H) = W^*\Omega(\alpha^*)\ [\ \overline{\mathbf{w}}(\frac{H}{H^*}) + \sum_{k=1}^{2K} c^{(k)}\mathbf{v}^{(k)}(\frac{H}{H^*})\ ], \qquad (4.1)$$

$$T(H) = T^{(0)} + T^*\ [\ \bar{t}(\frac{H}{\widetilde{H}^*}) + \sum_{k=1}^{K} \widetilde{c}^{(k)}\tau^{(k)}(\frac{H}{\widetilde{H}^*})\ ]. \qquad (4.2)$$

To simulate meteorological elements according to (4.1), (4.2), the values of dimensional reference scales and of transform coefficient (marked by superscript *) are simulated as standard distributions, while their statistical characteristics and all the other values are estimated on the basis of the measurement data. As demonstrated below, temperature values $T^{(0)}$ and $T^*$ are strongly correlated. So that the knowledge of $T^*$ provides for the value of $T^{(0)}$ as well.

In Fig. 8 (a-b) the measured longitudinal and transversal winds (the mean direction of wind $\alpha_m^*$ corresponds the longitudinal wind, according to Eq. (3.5)) and the temperature are compared with their approximations in the PCA transform. Note that the temperature curves are practically coincident in both Fig. 8 (c) and (f).

Thus, the dimensionless mean wind and temperature functions $\overline{\mathbf{w}}$ and $\bar{t}$ are computed by (3.6), the basic functions $\mathbf{v}^{(k)}$ and $\tau^{(k)}$ are calculated as the solution of corresponding eigenvalue problems, while reference scales $\{H^*, U^*, V^*, \widetilde{H}^*, T^*\}$ and the PCA transform coefficients $\{c^{(k)}\}$, $\{\widetilde{c}^{(k)}\}$ will be simulated by standard distributions with known statistical characteristics.

Since both wind and temperature are employed simultaneously in further applications, meteorological elements containing both wind and temperature are considered. So their statistical model is required taking into account correlation between them. In accordance with our basic idea of meteorological homogeneity (universality) of the data after scaling, we suppose that the wind-temperature correlation is fully described by the correlation between their characteristic reference scales $\{H^*, U^*, V^*, \widetilde{H}^*, T^*\}$. The former hypothesis adopted in the current study is that the following three ensembles are mutually independent: the ensemble of characteristic reference scales $\{H^*, U^*, V^*, \widetilde{H}^*, T^*\}$, the ensembles of the PCA transform coefficients for the wind $\{c^{(k)}\}$ and temperature $\{\widetilde{c}^{(k)}\}$. Statistical tests of this hypothesis were carried out and did not invalidate it for the main modes (from k=1,2 up to k=5,6 for wind basic functions $\{u^{(k)}, v^{(k)}\}$). Finally, the second hypothesis that we additionally assume is that the partial distributions of each of these three ensembles are normal. Statistical tests of the last hypothesis did not invalidate it.

Under these assumptions, it follows from Eqs. (3.13) that transform coefficients $\{c^{(k)}\}$ and $\{\widetilde{c}^{(k)}\}$ are independent for different *k* normally distributed random values with zero means and variances $\{\lambda^{(k)}\}$ and $\{\widetilde{\lambda}^{(k)}\}$. Hence, each of the transform coefficients may be simulated by the standard procedure of generating normally distributed random numbers. The simulation of coefficients $\{\widetilde{c}^{(k)}\}$ is performed analogously. After calculating the sample means and variances of the 5-dimensional vector of reference scales $\{H^*, U^*, V^*, \widetilde{H}^*, T^*\}$, this vector is simulated in the usual way as a normally distributed vector with known mean values and standard deviations (Table 1). Note that standard deviations of all the values in Table 1 except of velocity components are of the order of 10% of corresponding mean values. Standard deviations of the both-components of the perturbation velocity have the same order $20 \div 30$ *knots,* which are higher the corresponding mean values of *U*-component of velocity, and three times lower of its *V*-component. This is in compliance with the above-mentioned existence of the prevailing wind direction close to the direction of the *V*-component of the wind.

Quite realistic results of simulations carried out in accordance with the above-described procedure are presented in Fig. 8 (d-f) for wind, temperature and their approximations. Correlation coefficients between the reference scales are presented in Table 2. The reference scales are weakly correlated except for the reference temperature $T^*$ and wind component $V^*$ directed along the prevailing wind direction. The negative values of the correlation coefficients between two parameters indicate the opposite direction of their variation, e.g. rise in $T^*$ corresponds to decrease in $V^*$ under their joint variation.

Table 1. Means and standard deviations for dimensional reference scales of meteorological elements (over 3 years, jointly and separately)

|  |  | $H^*$, m | $U^*$, knot | $V^*$, knot | $\tilde{H}^*$, m | $T^*$, °C |
|---|---|---|---|---|---|---|
| 1997-99 | Mean | 12450 | -16.5 | 72.7 | 17445 | 88.7 |
|  | St. dev. | 1491 | 27.4 | 24.2 | 1323 | 8.4 |
| 1997 | Mean | 12472 | -15.3 | 70.0 | 17348 | 87.2 |
|  | St. dev. | 1478 | 23.7 | 24.0 | 1504 | 8.6 |
| 1998 | Mean | 12355 | -14.1 | 74.3 | 17585 | 88.2 |
|  | St. dev. | 1514 | 27.2 | 21.1 | 1285 | 7.7 |
| 1999 | Mean | 12469 | -20.1 | 74.3 | 17430 | 90.4 |
|  | St. dev. | 1501 | 30.8 | 26.8 | 1142 | 8.6 |

Table 2. Matrix of correlation coefficients for reference scales of meteorological elements

|  | $H^*$ | $U^*$ | $V^*$ | $\tilde{H}^*$ | $T^*$ | $T^{(0)}$ |
|---|---|---|---|---|---|---|
| $H^*$ | 1.0 | 0.04 | - 0.09 | 0.03 | 0.06 | 0.05 |
| $U^*$ | 0.04 | 1.0 | 0.09 | - 0.03 | -.17 | - 0.17 |
| $V^*$ | - 0.09 | 0.09 | 1.0 | 0.02 | -.45 | - 0.47 |
| $\tilde{H}^*$ | 0.03 | - 0.03 | 0.02 | 1.0 | -.14 | - 0.15 |
| $T^*$ | 0.06 | - 0.17 | - 0.45 | -0.14 | 1.0 | 0.93 |
| $T^{(0)}$ | 0.05 | - 0.17 | - 0.47 | -0.15 | 0.93 | 1.0 |

According to (4.2) the knowledge of characteristic parameter $T^{(0)}$ is needed for temperature simulation. Correlations with the characteristic parameter $T^{(0)}$ presented in Table 2 demonstrate their proximity to the all correlation coefficients for $T^*$, in particular, the extremely high correlation between the reference temperature $T^*$ and $T^{(0)}$. So the value of $T^{(0)}$ can be expressed with acceptable exactness through reference scale $T^*$.

## 5. Summary and Discussion

The joint statistical model of the local horizontal wind and temperature is developed on the basis of the standard on-site meteorological data set gathered in the operational weather service.

The model is developed for the practically interesting case of the strong winds (high-level jet streams), i.e. for $W_m^{max}>50$ *knot* when the altitude level of the maximum wind is highly expressed at $10 \cdot 10^3 \div 14 \cdot 10^3 \, m$ and varied about the mean value with a relatively low dispersion. Analysis of the meteorological data shows that about 70% of all winds observed are strong. In their turn strong winds have the west-sound prevailing direction (about 85% of their total number). The strong winds are accompanied by decrease of temperature with altitude up to the altitude level of the inverse temperature that occurs at $16 \cdot 10^3 \div 20 \cdot 10^3 \, m$.

To properly take into account the daily and seasonal variations of meteorological data the scaling procedure was developed. In particular, the appropriate dimensional reference scales were chosen and the scaling procedure was developed which reduce the measurement data for 2-dimensional wind vectors to a common base. The scaling procedure consists of the rotation by the reference angle along with the standard use of conversion scaling factors. A proper choice of the reference scales for a scaling procedure makes the measurement data meteorologically homogeneous and dimensionless. The Principal Component Analysis (PCA) is applied to the homogenized data for wind and temperature, which are considered as realizations of a random process with universal, i.e. independent of atmospheric conditions characteristics: means, PCA basic functions and parameters of standard distributions for the PCA transform coefficients. The features of the atmospheric conditions are fully described by statistical characteristics of the joint wind-temperature ensemble of dimensional reference scales. Since both wind and temperature are jointly employed in further applications, meteorological elements containing both wind and temperature are considered. So their joint statistical model is required taking into account the correlation between them. Additionally to the basis hypotheses of the present study, we have supposed the wind-temperature correlation is fully described by correlation between their dimensional reference scales. Statistical tests of the basic hypotheses were carried out and did not invalidate them from the practical point of view.

The mean wind speed and temperature are found to be near linear with altitude at least for altitudes lower than the altitude level of the maximum wind. Deviations from the mean wind and temperature are described by expansion over eigenfunctions of PCA transform. The model developed yields a fair approximation for observation data with a relatively small mean-square error that vanishes fast with growth of the mode number. Although any required exactness can be achieved, too high exactness has little meaning, and physically admissible values of the wind mean-square error correspond to 5- or even 3-mode approximation. Since the low-mode expansions are valid for the physically meaning approximation of the wind and temperature, the whole-world meteorological database with relatively rough vertical grid [1] may be used for the present analysis in other regions, where the high-level jets are prevailing. The dimensionless statistical characteristics of the local meteorology calculated independently for three consequent years are found to be very close to one another which demonstrates the annual stability of the meteorological database at least for the main statistical characteristics of the low order modes. The annual characteristics of the higher order modes, however, were subject to significant variations from year to year. The high effectiveness of the proposed approach is provided by a similarity of wind – temperature vertical profiles, which allow to carry out the statistical modeling in the low-dimension space of the dimensional reference scales for the wind, temperature and altitude levels. In conclusion, note that the proposed approach can be applied to similar data independently from used measurement' apparatus and location.

---

[1] **www.cdc.noaa.gov/cdc/data.ncep.reanalysis.html**